# Unique dynamic crossover in supercooled x,3-dihydroxypropyl acrylate (x = 1, 2) isomers mixture

Szymon Starzonek[1♦], Aleksandra Kędzierska-Sar[1,2], Aleksandra Drozd-Rzoska[1♦], Mikołaj Szafran[2], Sylwester J. Rzoska[1♦]

[1]*Institute of High Pressure Physics, Polish Academy of Sciences, ul. Sokołowska 29/37, 01-142 Warsaw, Poland*

[2]*Department of Chemical Technology, Warsaw University of Technology, ul. Noakowskiego 3, 00-664 Warsaw, Poland*

[♦]*Warsaw Dielectrics Group*

**Abstract**

The previtreous dynamics in glass forming monomer, glycerol monoacrylate (GMA), using broadband dielectric spectroscopy (BDS) was tested. Measurements revealed the clear dynamic crossover at temperature $T_B = 254$ K and the time scale $\tau(T_B) = 5.4$ ns for the primary (structural) relaxation time and no hallmarks for the crossover for the DC electric conductivity $\sigma_{DC}$. This result was revealed via the derivative-based and distortions-sensitive analysis $d\ln H_a/d(1/T)$ vs. $1/T$, where Ha is for the apparent activation energy. Subsequent tests of the fractional Debye-Stokes-Einsten relation $\sigma_{DC}(\tau_\alpha)^S = const$ showed that the crossover is associated with $S = 1$ (for $T > T_B$) $\rightarrow S = 0.84$ (for $T < T_B$). The crossover is associated with the emergence of the secondary beta relaxation which smoothly develops deeply into the solid amorphous phase below the glass temperature $T_g$.

**Introduction**

One of the most mysterious forms of the solidification is the vitrification at the glass temperature when passing from the metastable ultraviscous liquid/fluid to the metastable amorphous solid glass state [1]. Although the glass temperature $T_g$ cannot be directly considered as the phase transition, due to its stretched and depending on the cooling rate nature, it is associated with far previtreous effects signaling the vitrification even 100 K above $T_g$. This inherent feature allows for the 'remote' estimation of $T_g$ value. Regarding the previtreous 'dynamic effects' one can recall such 'universal' behavior as (i) the non-Arrhenius evolution on the primary relaxation time ($\tau$), viscosity ($\eta$) or electric conductivity ($\sigma$), (ii) the non-Debye distribution of relaxation time, (iii) the dynamic crossover between the ergodic and non-ergodic dynamical regimes, most often associated with the time scale $\tau_B = 10^{-7\pm1}$ s, (iv) the emergence of the secondary relaxation in the ultraslowing domain for $T < T_B = T(\tau = \tau_B)$. The glass temperature is by convention associated with the time scale of the primary relaxation process, $\tau(T_g) = 100$ s which corresponds to empirically observed $T_g$ value in the heat capacity scan for the most standard cooling rate 10 K·min⁻¹ [2, 3]. The fascinating previtreous 'universal' behavior of dynamic properties on approaching the glass temperature is undoubtedly one of key reason that the problem of in-deep fundamental understanding of the glass transition is indicated as one of the greatest challenges for 21st condensed matter physics, with enormous importance for the material engineering implementations [1-4]. One of lacking issues in this domain, which may appear essential for the ultimate insight, are studies covering domains on both sides of the glass transition, i.e. the metastable ultraviscous liquid and the metastable amorphous solid. In fact conclusive research on the latter is particularly difficult due to the fact that when passing the time scale $\tau(T_g)$ the system time scale by decades exceeds the experimental one, making ultimate conclusions puzzling [2, 3].

In this report it is shown that the secondary relaxation process [1-3], several decades faster than the primary relaxation time, seems to continue smoothly deeply within the solid glass state. Its evolution in this solid state clearly correlates with such basic feature of the liquid state as the dynamic crossover. Studies were carried out in glycerol monoacrylate monomeric system, important for application in ceramics formation.

## I. Experimental

The tested glass former was glycerol monoacrylate which is a low-toxic monomer used for manufacturing ceramic materials through the gelcasting process, what makes results of this paper additionally important for developing in situ monitoring of this process. Gelcasting is near-net-shaping process, which minimizes machining of the green body [5-7]. The monomer, present in the ceramic slurry, polymerizes during the forming process inside the mold.

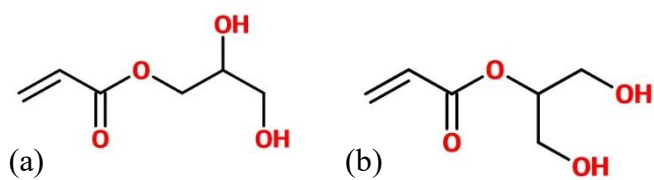

***Fig.1.*** *Molecular structures of 2,3-dihydroxypropyl acrylate (a) and 1,3-dihydroxypropyl acrylate (b).*

As a result, the obtained green body models the shape of the mold. Glycerol monoacrylate is a mixture of two isomers: 2,3-dihydroxypropyl acrylate and 1,3-dihydroxypropyl acrylate. It has a number of advantages when compared to different commercial monomers. It is water soluble, which allows conducting the process in water. Due to the presence of two –OH groups, the use of a crosslinking agent is not necessary. The obtained green bodies have high rigidity when compared to those obtained from commercial monomers with additions of crosslinking agents [5-7]. The glycerol monoacrylate (GMA) mixture synthesized at the Warsaw University of Technology was used. The tested sample was composed from: 2,3-dihydroxypropyl acrylate (70 %wt.) and 1,3-dihydroxypropyl acrylate (30 %wt.). Their molecular structures are shown in Fig. 1.

The complex dielectric permittivity $\varepsilon^*(f) = \varepsilon'(f) - i\varepsilon''(f)$ and electric conductivity $\sigma(f)$ was measured using the Alpha-A impedance analyzer (Novocontrol) with 6-digits resolution at ambient pressure ($P = 0.1$ MPa) over a frequency range from $10^7$ Hz to 1 Hz. During measurements the sample was maintained under the nitrogen gas flow at a temperature range between 273 K and 143 K. The temperature was controlled using Quatro Cryosystem (Novocontrol) with stability better than $\Delta T = 0.1$ K. The liquid GMA was put into two round parallel plates measurement capacitor made from steel with diameter $2r = 20$ mm. Teflon® ring was used as a spacer yielding a macroscopic gap $d = 0.2$ mm. The dielectric loss spectrum gives a background for relaxation times analysis $\tau = (2\pi f)^{-1}$, where $f$ is a frequency of peak's maximum; and DC-conductivity $\sigma_{DC}$. All dielectric spectra were fitted using Havrilliak-Negami function [1-3].

## II. The evolution of relaxation times and the dynamic crossover

One of key features of the glass previtreous dynamics is the super-Arrhenius behavior of the primary (α, alpha) relaxation time [1-4]:

$$\tau(T) = \tau_0 \exp\left(\frac{E_A(T)}{RT}\right) \qquad (1)$$

where $E_A(T)$ stands for the apparent activation energy. This relation converts into the classical Arrhenius dependence when in the given temperature domain $E_A(T) = E_A = const$.

The generally unknown form of the evolution of the apparent activation energy causes in the portrayal of experimental data one have to use ersatz dependences. For decades the most popular was the Vogel-Fulcher-Tammann (VFT) dependence [1-4, 8]:

$$\tau(T) = \tau_0 \exp\left(\frac{D_T T_0}{T-T_0}\right), \qquad (2)$$

where $D_T$ denotes the fragility strength coefficient, $T_0$ is for the ideal glass 'transition' temperature hidden in the solid glassy state and $R$ stands for the gas constant. By comparing Eqs. (1) and (2) for the apparent activation energy following equation has been obtained $E_A(T) = RD_T/(T_0^{-1} - T^{-1})$.

In recent years, novel equations yielding more optimal $\tau_\alpha(T)$ or $\eta(T)$ parameterizations and questioning the general validity of the VFT relation appeared [9-16]. Particularly, successful appears the relation empirically introduced by Waterton [17] in 1932 and recently validated as the output of the constraint theory applied to the Adam-Gibbs model by Mauro et al. [11], namely:

$$\tau(T) = \tau_0 \exp\left[\frac{K}{T}\exp\left(\frac{C}{T}\right)\right]. \qquad (3)$$

It is notable that it has no final temperature divergence, which is the characteristic feature of the VFT relation. Notwithstanding, the ultimate parameterization of the temperature dependence of the primary relaxation or viscosity in the ultraviscous domain remains the puzzling issue. Worth recalling here is the recent 'model-free' ('fitting-free') analysis based on the apparent activation temperature index, which clearly shows the limited fundamental validity of all basic equation used so far, including the VFT and WM (MYEGA) ones.

One of 'universal' of the previtreous dynamics is the change in the form of the super-Arrhenius (SA) behavior most often occurring at the 'magic' time scale $\tau(T_B) \sim 0.1$ μs, although there is also a notable number of exceptions [2, 18]. The key tool for the detection of the crossover is the plot proposed two decades ago by Stickel et al. [19]: $\phi_T(T) = [d\log_{10}\tau_\alpha(T)/d(1/T)]^{-0.5}$ vs. $1/T$. As shown in ref. [9, 10] this plot is related to the evolution of the apparent activation enthalpy with the underlying background of the hypothetical general validity of the VFT equation, namely:

$$\phi_T(T) = (H_A(T))^{-0.5} = A - \frac{B}{T} \qquad (4)$$

Where the apparent activation energy: $H_A(T) = R\frac{d\ln\tau_\alpha(T)}{d(1/T)}$. Note that $\ln\tau_\alpha = \log_{10}\tau_\alpha / \log_{10} e$. The 'linearization-focused' transformation of experimental data via eq. (4) enable the unequivocal estimation of basic parameters in the VFT equation: $T_0 = |B/A|$, $D_T = 1/|AB|$.

Regarding the significance of the dynamic crossover phenomenon worth recalling is the statement from ref. [20]: '*…one may expect that understanding the meaning of the dynamic crossover phenomenon may be essential for the ultimate understanding of the puzzling nature of the glass transition. …$T_B$ appears to be more relevant than $T_g$ or $T_0$…*'.

It is notable that the estimation of the location of $T_B$ via the 'Stickel plot' or eq. (4) assume the fundamental validity of the VFT description for the super-Arrhenius evolution of the primary relaxation time or related properties. However, as shown in ref. [14-16] this is the case only for a limited number of glass formers. Consequently, the question arises for the 'fitting-free' way determining of the dynamical crossover. In Refs. [14-16] it was noted that the domain of validity of the MYEGA eq. (3) can be estimated via the following derivative based and distortions – sensitive plot based on $\tau_\alpha(T)$ experimental data: $\ln[H_A'/(1 + C/T)]$ vs. $1/T$ [21]. The linear plot indicates the domain validity of such description and the subsequent linear regression fit yield optimal values of basic parameters in Eq. (3). Following this result in Refs. [9, 10] the new 'model-free'/'fitting-free' way of analysis indicated changes in the form of description of $\tau_\alpha(T)$ behavior was proposed and tested for several glassforming systems: $\ln(H_A(T)/R)$ vs. $1/T$, where the apparent activation enthalpy $H_A(T) = d\ln\tau_\alpha/d(1/T)$.

Another possibility of the 'fitting-free' is the analysis focused on the emergence of the orientational-translational decoupling in the non-ergodic domain below $T_B$ [2, 3]. For BDS related experimental data it is expressed via the fractional Debye-Stokes-Einstein relation:

$$(\sigma_{DC}(T))(\tau_\alpha(T))^S = const \; [22]. \tag{5}$$

The value of the exponent $S < 1$ is expected for $T_g < T < T_B$ and indicated the orientational-transitional decoupling, i.e. the speed up of orientational processes in comparison to translational ones. For $T > T_B$ the exponent $S = 1$ and both processes are coupled, i.e. the time scale of orientational and translational dynamics are paralleled coupling appears. Most liquids are characterized by $0.7 < S < 1$ [22]. However, in some liquid crystal's phases the exponent can be even lesser than 0.5 [23, 24]. In Ref. [22] the link of the exponent $S$ activation enthalpy of the process was shown.

One of 'universal' features of the previtreous dynamic is the emergence of the secondary, 'beta' ($\beta$), relaxation process for $T < T_B$. Both 'alpha' and 'beta' relaxation processes splits at $T = T_B$ and when reaching $T_g$ the difference between related time scale reaches several decades. It is notable that the secondary relaxation most often follows the Arrhenius pattern. In the case of polymeric glass formers the beta relaxation is linked to internal molecular relaxation processes [2, 3].

## III. Results and discussion

Fig. 2 presents examples of dielectric loss spectra $\varepsilon''(f)$ for supercooled GMA samples when decreasing temperatures from 273 K to 193 K. The emergence of weak primary (structural) relaxation $\tau_\alpha$, deep-glassy process $\beta$ and strong conductivity contribution are visible. The characteristic slop change of the low frequency part of spectra at the $\log_{10} \varepsilon''$ at $\log_{10} f$ plot is related to electrode polarization appearing in highly conductive systems, to which GMA belongs. This low frequency part of spectra also serves for determining the DC-conductivity. Regarding the basic structural α-process dielectric loss curves exhibit a typical for complex glassy dynamics non-Debye shape. The obtained evolution of relaxation times is presented in Fig. 3a, covering both the ultraviscous liquid and solid states. Fig. 3b shows results of the supplementary distortions-sensitive and derivative based analysis of data, following the method recalled in the previous section.

***Fig.2.*** *Dielectric loss spectra $\varepsilon''(f)$ for glycerol monoacrylate mixture at the decreasing temperature from 273 K down to 193 K. The two relaxation processes: structural (α) and glassy (β); and conductivity contribution $\sigma_{DC}$ are presented.*

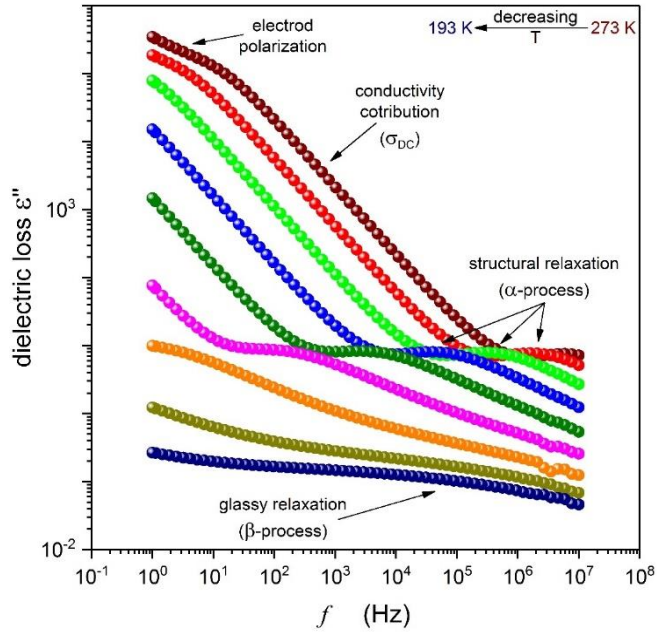

The horizontal solid line (red) indicates the behavior well portrayed by the simple Arrhenius relation Eq. (1). It is associated with the activation energy $E_A = 27$ kJ·mol$^{-1}$. On cooling there is almost 100 K broad Super-Arrhenius domain, which can be well portrayed by the VFT equation (2): $T_0 = 159.10$ K, $D_T = 10.15$. Using the formula $m = m_{P=0.1\,\text{MPa}} \approx 16 + 590/D_T$, which is related to the fragility coefficient $m = 76$, for $m = \left[d\log_{10}\tau_\alpha/d(T_g/T)\right]_{T\to T_g}$ [2, 3]. Notable is the very clear appearance of the dynamic crossover at $T_B = 254$ K, for the relaxation time $\tau_\alpha(T_B) = 5.4$ ns.

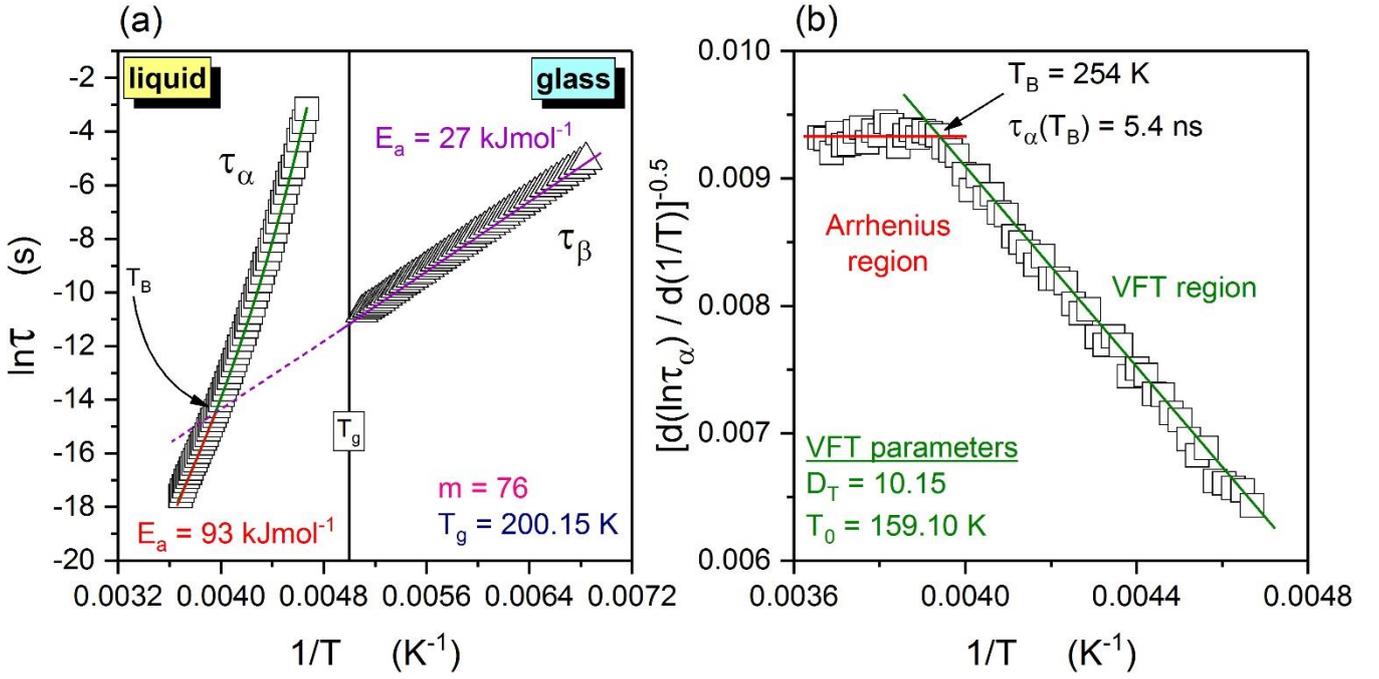

***Fig.3.*** *(a) Distribution of relaxation times: primary (α) and secondary (β) for liquid and glassy glycerol monoacrylate mixture. The solid vertical line presents the glass transition temperature $T_g$ = 200.15 K. From the Arrhenius behaviors of relaxation times activation energies were calculated. (b) Shows derivative-based analysis of the structural relaxation time (α), so called 'Stickel plot'. This linearization allows to obtain dynamic regions decreasing temperature and calculate VFT parameters, which can be used for estimating the fragility index m = 76. The dynamic crossover occurs at a temperature $T_B$ = 254 K and the relaxation time in this point $\tau_\alpha(T_B)$ = 5.4 ns.*

Translational processes are characterized within BDS spectrum by the DC electric conductivity. Its temperature evolution within the ultraviscous liquid state is shown in Fig. 4a, revealing a clear Super-Arrhenius behavior. The distortions sensitive and derivative based plot $d\ln\sigma_{DC}^{-1}/d(1/T)$ vs. $1/T$ shows no hallmarks of the translational-orientational decoupling for the electric conductivity when passing $T_B$, clearly manifesting in $\tau_\alpha(T)$ evolution. The dynamic crossover is present, while testing the orientational processes and absent for translational ones. The evolution of $\sigma_{DC}^{-1}(T)$ is well portrayed by the VFT relation (the solid curve in Fig. 4a and it is associated with the fragility index $m_\sigma = 64.45$.

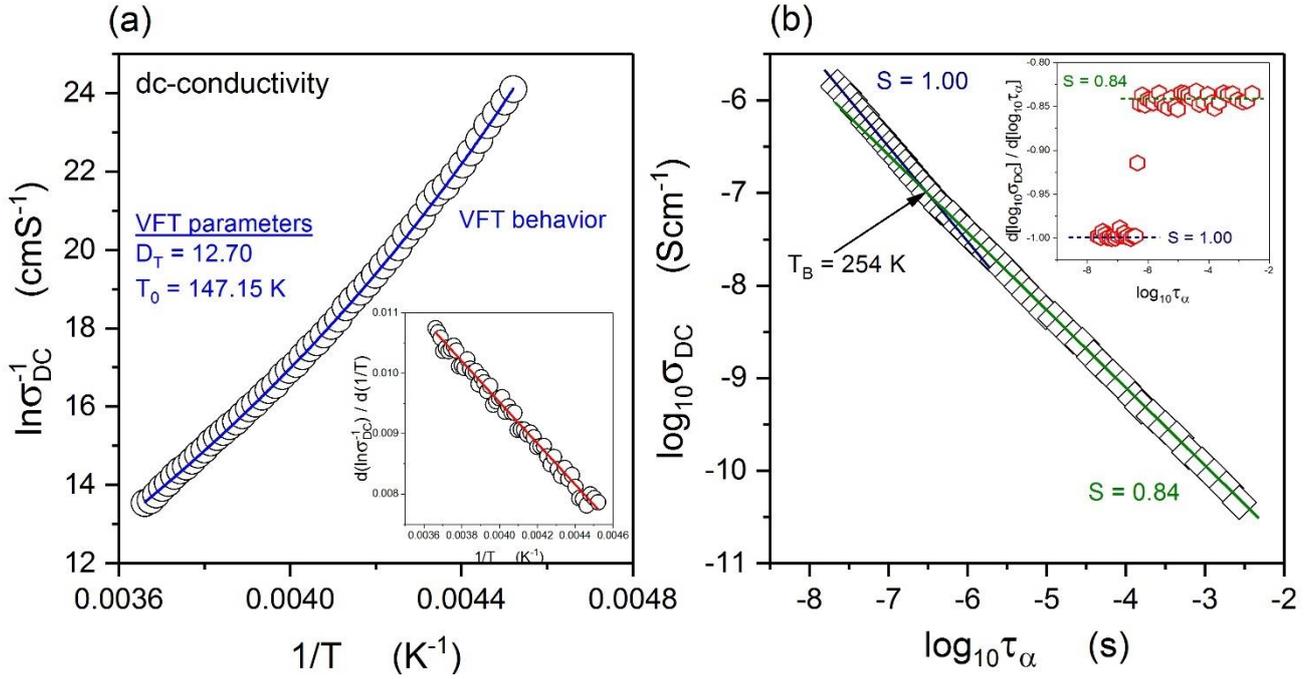

***Fig.4.*** *(**a**) The VFT behavior of the dc-conductivity reciprocal. (**b**) The fractional Debye-Stokes-Einstein test. At lower temperatures the transitional-orientational decoupling occurs. The inset shows derivative analysis with two characteristics values of the FDSE exponent, which are different slopes: S = 1.0 for the Super-Arrhenius region, and S = 0.84 for the VFT domain*

The translational-orientational (TO) coupling/decoupling is directly tested in Fig. 4 (b) via the test of the validity of the Debye-Stokes-Einstein. The inset shows derivative of $d \log_{10} \sigma_{DC} / d \log_{10} \tau_\alpha$ of data from the main part of the plot, conforming the occurrence of the dynamic crossover at the well-defined temperature $T_B$, the same as in Fig. 3. The crossover is associated with the shift from the region of obeying the DSE law and the translational-orientational coupling ($S = 1$) to the decoupling region ($S = 0.84$) in the immediate vicinity of $T_g$. It was derived in ref. [22] that the fractional DSE exponent:

$$S = \frac{H_A^{\sigma_{DC}}}{H_A^{\tau_\alpha}} = \frac{m^{\sigma_{DC}}}{m^{\tau_\alpha}}, \quad \text{(for } P = \text{const)} \tag{6}$$

where $H_A^{\sigma_{DC}}$, $H_A^{\tau_\alpha}$ are apparent activation enthalpies and $m^{\sigma_{DC}}$, $m^{\tau_\alpha}$ are 'fragility indexes' calculated from DC-conductivity and structural relaxation data respectively. For glycerol monoacrylate (GMA) the exponent $S = 0.84$ is in fair agreements with the value $S = m^{\sigma_{DC}}/m^{\tau_\alpha} = 64.45/76 \approx 0.848$.

The common, 'universal' feature of dynamics in the non-ergodic domain closed to $T_g$ is the emergence of secondary relaxation process. This is also the case of GMA dynamics. However, this process becomes particularly well visible in the solid amorphous state, for $T < T_g$. The detection of this beta process in the solid amorphous phase is easily possible even 50 K below $T_g$ where $\tau_\beta(T_g - 50\text{ K}) \sim 10^{-4}$ s, i.e. it is experimentally detectable. The direct detection of the structural relaxation time $\tau_\alpha$ for such temperature is in practice not possible, because it approximately exceeds millions of years value.

## IV. Conclusions

This report present the analysis of the dynamics in glycerol monoacrylate (GMA) mixture, being the glass former important in gelcasting technology in ceramics. The tested system shows relatively high electric conductivity, but despite this difficulty the broadband dielectric scanning enabled tests of the temperature evolution of orientational and translational relaxation processes. The structural relaxation time $\tau_\alpha(T)$ shows a clear dynamic crossover from the Arrhenius to Super-Arrhenius behavior at the temperature $T_B \approx 254$ K. The crossover is absent for the DC electric conductivity, related to translational processes. The crossover is clearly visible also in tests focused on the occurrence of the fractional Debye-Stokes-Einstein law. In the solid amorphous state the primary relaxation time $\tau_\alpha(T)$ ceases to be directly detectable due to the enormous increasing of the system time scale. Notwithstanding, there is a clear manifestation of the qualitatively faster relaxation process, which after the extrapolation to the ultraviscous liquid state merges with the structural relaxation time at $T = T_B$, what allows to claim that for the given system the secondary relaxation processes smoothly develops from the metastable ultraviscous liquid state to the metastable solid glass state without a hallmark when passing the solidification point at the glass temperature.

## V.  Acknowledgements

For some of the authors (SS, SJR, ADR) the research was supported by the NCN OPUS (Poland) Project ref. 2016/21/B/ST3/02203 (UMO-2016/21/B/ST3/02203). For AKS this work was financed from National Science Centre in Poland from grant number 2014/13/N/ST5/03438.